\documentclass[aps, prl, twocolumn, showpacs, floatfix,10pt,superscriptaddress]{revtex4-2}
\usepackage{amsmath}
\usepackage{amsfonts}
\usepackage{amsbsy}
\usepackage{amssymb}
\usepackage{graphicx}
\usepackage{textcomp}
\usepackage[caption=false,singlelinecheck=false]{subfig}
\usepackage{xcolor}
\usepackage{mathrsfs}
\usepackage{mathtools}
\usepackage{bm}
\usepackage{gensymb}
\usepackage{braket,wasysym}
\usepackage[colorlinks,linkcolor=blue,citecolor=red,filecolor=magenta,urlcolor=red,breaklinks]{hyperref}
\usepackage[bottom]{footmisc}
\usepackage{verbatim}

\hypersetup{colorlinks=true, urlcolor=blue, citecolor=red, pdfborder={0 0 0}}
\usepackage{breakurl}
\usepackage{natbib}

\allowdisplaybreaks

\newcommand{\tcm}[1]{\textcolor{magenta}{#1}}

\normalfont
\def\be{\begin{equation}}
	\def\ee{\end{equation}}
\def\bea{\begin{eqnarray}}
	\def\eea{\end{eqnarray}}

\begin{document}
\title{Immortal quantum correlation in quasiperiodic quasi-1D system}

\author{Junmo Jeon}
\email{junmo1996@kaist.ac.kr}
\affiliation{Korea Advanced Institute of Science and  Technology, Daejeon 34141, South Korea}
\author{SungBin Lee}
\email{sungbin@kaist.ac.kr}
\affiliation{Korea Advanced Institute of Science and  Technology, Daejeon 34141, South Korea}

\date{\today}
\begin{abstract}
The prevailing view on long-range correlations is that they typically attenuate uniformly with distance and temperature, as most interactions either exhibit short-range dominance or decay following a power law. In contrast to this belief, this study demonstrates that the intricate interplay between quasiperiodicity and the quasi-1D nature of subbands can result in strong long-range coupling without attenuation, a phenomenon referred to as an \textit{immortal interaction}. 
Exemplifying a periodically stacked Fibonacci chain, we uncover an \textit{immortal interaction} with greatly enhanced, persistent long-range coupling. 
Using negativity, it is shown that this interaction creates stable entanglement that endures over long distances and remains robust at finite temperatures. Additionally, unconventional logarithmic scaling entanglement is revealed, deviating from the traditional area law. These findings offer quasiperiodic quasi-1D systems as a novel platform for sustaining stable entanglement across exceptionally long distances, even in the presence of finite temperature effects.
%we show this immortal interaction creates stable entanglement that endures over long distances and remains robust at finite temperatures. Furthermore, we reveal unconventional logarithmic scaling entanglement, different from the traditional area law. Our findings unveil quasiperiodic quasi-1D systems as a novel platform for sustaining stable entanglement across exceptionally long distances, even in the presence of finite temperature effects.
\end{abstract}
\maketitle

\textit{\tcm{Introduction---}}
Entanglement, a fundamental quantum correlation, has fueled a broad spectrum of applications that are central to the second quantum revolution\cite{lars2018second,hu2023progress}. 
Over the past century, entanglement has propelled advancements across various scientific domains, including quantum computing\cite{cai2015entanglement}, condensed matter physics\cite{zeng2015quantum,cramer2011measuring,hamma2008entanglement}, and optics\cite{togan2010quantum,fiaschi2021optomechanical}. One of the most significant research emerging from the discovery of entanglement is quantum information processing. A prime example is quantum teleportation — a protocol that enables the transfer of an unknown quantum state using preexisting entanglement and a classical communication channel\cite{yin2012quantum}. The effectiveness of such applications critically depends on the availability of long-range entanglement between the sending and receiving parties\cite{agarwal2023long}.

In condensed matter systems, long-range entanglement also carries significant importance. By analyzing their scaling behavior, it can distinguish different phases of matters and topological orders in the presence of strong correlation, such as quantum spin liquids\cite{osterloh2002scaling,savary2016quantum,isakov2011topological,lee2013entanglement}. 
Moreover, recent studies explore various experimental probes capable of measuring quantum entanglement in multipartite systems, with neutron scattering experiments being one method for extracting quantum Fisher information as an example\cite{sabharwal2024witnessing,scheie2021witnessing,gabbrielli2018multipartite}.
Accordingly, both theoretical and experimental researchers are highly invested in preparation and control of long-range entanglement in condensed matter systems\cite{gong2017entanglement,defenu2023long}. %\tcr{ref : entanglement related review paper, entanglement spectrum entropy .. cite many refs here }
In typical condensed matter systems, however, the entanglement between distant particles tends to decay uniformly since the interactions responsible for generating entanglement are generally short-range or follow a power-law decay, diminishing to negligible levels over experimentally accessible distances\cite{defenu2023long,degli2006long,jin2004localizable}. Hence, long-range entanglement is usually obscured by thermalization\cite{gabbrielli2018multipartite}. Thus, to effectively utilize long-range entanglement and explore strongly correlated physics, extremely low temperatures are necessary\cite{heiss2008fundamentals}. This requirement presents substantial challenges for exploring various applications of quantum correlations, including long-distance quantum teleportation and long-range entangled phases of matter.

Some slowly decaying long-range interactions like unscreened Coulomb interaction, have been expected as key drivers of long-distance entanglement in many-body quantum systems\cite{eisert2008area,defenu2023long,eisert2006general}. Such interactions can accumulate mutual information across a subsystem, leading to long-range entanglement which diverges logarithmically or follows a volume law, even in non-critical systems\cite{fannes2003entropy}. Various experimental platforms, such as cold atoms\cite{arguello2022tuning}, Rydberg atom arrays\cite{saffman2010quantum}, superconducting islands\cite{estrada2024correlation} and dipolar systems\cite{lahaye2009physics} have been explored for their ability to fine-tune interactions and enhance entanglement over long distances\cite{islam2013emergence}. Utilizing these systems, however, is challenging since diminished interaction strength over long distances requires a large number of qubits to maintain coherence while serving as intermediaries between widely separated target qubits\cite{eisert2008area,liao2022end}. This makes these systems prone to environmental disturbances like decoherence, limiting their practical use for long-range entanglement\cite{jacquod2009decoherence,buchleitner2008entanglement}. Hence, finding a platform that can achieve stable long-range entanglement over practical distances despite environmental influences remains an unsolved task.

\begin{figure}
  \includegraphics[width=0.48\textwidth]{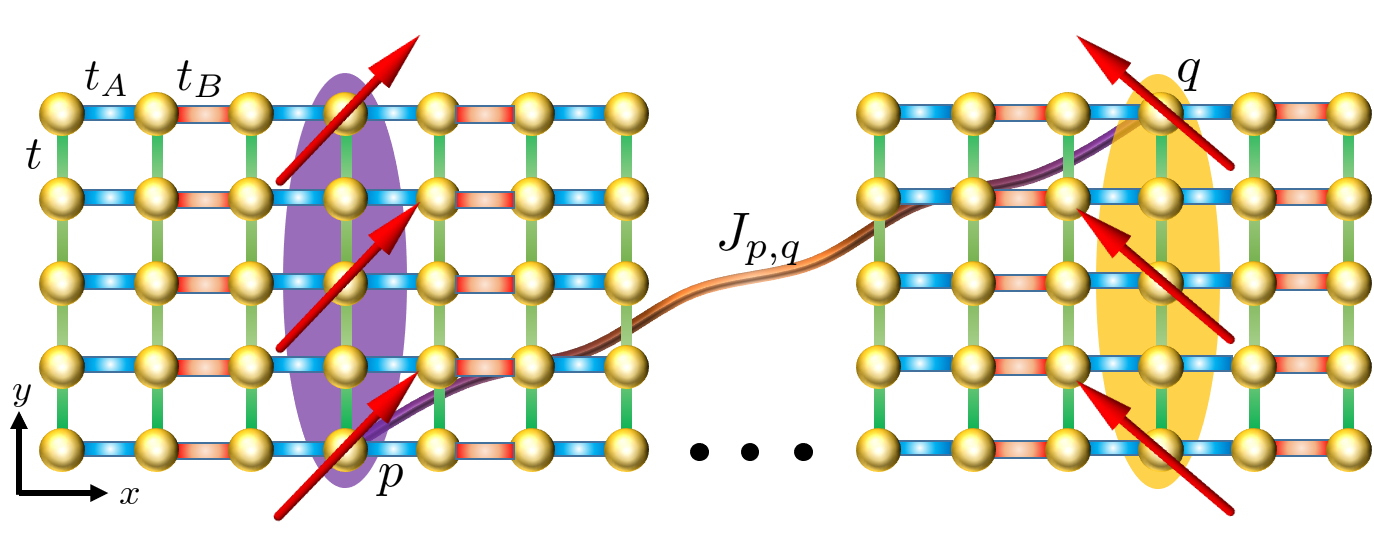}
  \caption{Schematic illustration of the quasiperiodic quasi-1D system, consisting of periodically stacked quasiperiodic chains. Yellow balls represent the atomic sites. Red and blue bars represent different hopping integrals, $t_A$ and $t_B$ in each quasiperiodic chain along the $x$ direction, respectively. Green bars represent uniform hopping integral, $t$ along the $y$ direction. Red arrows represent localized spin moments that interact locally with the itinerant electron spin. Itinerant electrons in such quasiperiodic system mediate indirect long-range interaction between two localized spins at the $p$ and $q$ sites, say $J_{p,q}$. As $t_A$ deviates from $t_B$, $J_{p,q}$ is enhanced, even when the $p$ and $q$ sites are in different layers and are separated by a distance on the order of $10^3$ sites. Violet and orange shaded regions are drawn for emphasizing two widely separated parties of the localized spins. Each party contains $N=(L_y+1)/2$ localized spins, where $L_y$ is the number of quasiperiodic chains.(See the main text for details)}
  \label{fig: system}
\end{figure}

In this work, we study the quasiperiodic quasi-one-dimensional system, exemplified by a periodically stacked Fibonacci chain, and explore anomalously persistent long-range interactions and their stable quantum correlations. This system exhibits quasiperiodicity along the $x$ direction and periodicity along the orthogonal $y$ direction (see Fig. \ref{fig: system}).
%\tcg{In this letter, building on our previous work, we investigate strong long-range interactions and entanglement characteristics in a periodically stacked quasiperiodic chain system, which features both quasiperiodicity and periodicity along orthogonal directions, denoted as $x$ and $y$, respectively (See Fig.\ref{fig: system}).}
%\tcr{it sounds like extension of our previous work. Instead, emphasize what is the novel feature of current work, answering to the question raised in the previous paragraph?}
Notably, the indirect spin interaction in this system is not only anomalously enhanced over long distances in the quasiperiodic $x$-direction but, even more strikingly, it exhibits no attenuation along the periodic $y$-direction — an effect we refer to as the  \textit{immortal interaction}. This unique phenomenon stems from the interplay of quasiperiodicity and quasi-1D nature of the system, where critical, non-Bloch type wave functions of each quasiperiodic chain form subbands because of the stacking. These intertwined quantum effects give rise to unusually strong interactions that are exceptionally persistent and non-decaying. By employing entanglement negativity, we show that this interaction leads to very stable entanglement, even over long distances and with finite temperature.
Furthermore, the immortal interaction enforces entanglement monogamy, leading to an unconventional logarithmic scaling contrast to the conventional area law. Our work presents a novel platform that maintains stable entanglement across vast distances and under finite temperature conditions, highlighting its unusual scaling properties.

\textit{\tcm{Periodically stacked quasiperiodic chains---}} Let us consider identical quasiperiodic chains that are periodically stacked as shown in Fig.\ref{fig: system}. We assume that the number of quasiperiodic chains, $L_y$,  is much smaller than the length of each quasiperiodic chain, $L_x$, thus a quasiperiodic quasi-1d system. To provide a concrete example, we consider the 1D Fibonacci quasicrystal as a specific instance of a quasiperiodic chain. 
However, we emphasize that our results are independent of the specific choice of the quasiperiodic chain.
In the Fibonacci case, the system consists of two types of nearest-neighbor atomic distances, A and B, each corresponding to different hopping integrals, denoted  $t_\mathrm{A}$ and $t_\mathrm{B}$, respectively\cite{jagannathan2021fibonacci}. The strength of quasiperiodicity is given by their ratio, $\kappa=|\log(t_\mathrm{A}/t_\mathrm{B})|$. We assume $t_\mathrm{A}\le t_\mathrm{B}$. The infinitely long Fibonacci quasicrystal is generated by a successive substitution map, $\mathrm{A}\to \mathrm{AB}$ and $\mathrm{B}\to\mathrm{A}$\cite{jagannathan2021fibonacci}. For clarity, we refer to the $x$ direction as quasiperiodic and the $y$ direction as periodic.

Let us consider the nearest neighbor tight binding Hamiltonian Eq.\eqref{H} for the electronic state, along with localized electrons positioned at a specific set of lattice sites, $\mathcal{D}$.
\begin{align}
\label{H}
&H=T_x+T_y+J_K\sum_{p\in\mathcal{D}}\vec{S}_p\cdot\vec{s}_p.
\end{align}
Here, $T_x$ and $T_y$ represent the nearest neighbor tight-binding Hamiltonian along $x$ direction and $y$ direction, respectively. In detail, $T_x=-\sum_{i,j}(t_{i,i+1}c_{i+1,j}^\dagger c_{i,j}+h.c.)$ and $T_y=-t\sum_{i,j}(c_{i,j+1}^\dagger c_{i,j}+h.c.)$, where $c_{i,j}^\dagger$ and $c_{i,j}$ are electron creation and annihilation operators for the site $(i,j)$. Note that $T_x+T_y$ has sublattice symmetry with open boundary conditions we consider here. 
Given that the hopping integral along the $y$ direction is $t=1$, the average hopping integral along the 
$x$-direction is also set to be $1$. $\vec{S}_p$ and $\vec{s}_p$ represent the localized electron spin and the itinerant electron spin at the $p$ site, respectively. $J_K<0$ is the local exchange coupling between the localized and itinerant electron spins, which is perturbatively small compared to the average hopping integrals, $|J_K| \ll t$. By integrating out the electronic degrees of freedom, one can get the long-range indirect exchange coupling between different localized spins, $\vec{S}_p$ and $\vec{S}_q$, $J_{p,q}$, given by,
\begin{align}
\label{RKKY2}
J_{p,q}=\frac{J_K^2}{4}\sum_{m,n,E_m\neq E_n}&\mathcal{R}[\psi_m(p)\psi_m(q)^*\psi_n(q)\psi_n(p)^*] \\ &\times \frac{n_F(E_n-E_F)-n_F(E_m-E_F)}{E_n-E_m}.\nonumber
\end{align}
Here, $\hbar=k_B=1$, $E_F$ is the Fermi level. $\mathcal{R}[x]$ is real part of $x$, respectively. $n,m$ are indices of the eigenstates, and $\psi_{n}(p)=\langle p \vert n\rangle$ is the wave function of the energy eigenstate $\vert{n}\rangle$ whose energy is $E_n$ at the $p$ site. $n_F(x)=(1+\exp(\beta x))^{-1}$ is the Fermi distribution function, where $\beta=1/\tau$ and $\tau$ is temperature\cite{jeon2023anomalous}. Hence, $J_{p,q}$ in Eq.\eqref{RKKY2} has temperature dependence. In conventional crystals, the magnitude of $J_{p,q}$ decays according to a power-law, especially in 2D systems, $|J_{p,q}|\sim R_{p,q}^{-2}$, where $R_{p,q}$ is the distance between $p$ and $q$ sites\cite{chaikin1995principles}.

The energy levels near the Fermi level play a crucial role in the long-range coupling described by Eq.\eqref{RKKY2}\cite{chaikin1995principles}. Specifically, when states near the Fermi level are not highly extended but instead exhibit critical behavior, the coupling strength between certain sites gets enhanced\cite{jeon2023anomalous}. This effect is further amplified by the strong quasiperiodicity, which makes critical states concentrated on a few widely separated sites, leading to an anomalous enhancement of the coupling strength between many pairs of sites that are far apart within each quasiperiodic layer\cite{jeon2023anomalous}. Such strongly interacting sites can be identified in each quasiperiodic chain by their local tiling patterns\cite{jeon2023anomalous}. Furthermore, since quasiperiodicity discretizes the electronic spectrum\cite{jagannathan2021fibonacci}, only a portion of subbands originating from the periodic stacking would be dominant in Eq.\eqref{RKKY2}. This leads to an anomalously strong long-range coupling that does not attenuate along the stacking direction, as we will discuss further.

%since the quasiperiodicity discretizes the electronic spectrum, only a portion of the subbands originating from the periodic stacking would be dominant in Eq.\eqref{RKKY2}. This gives rise to the anomalous strong long-range coupling which does not attenuating along the stacking direction as we will now discuss.

%\tcg{Here, we focus on the half-filled case with $E_F=0$ in the presence of the sublattice symmetry. Based on our previous work on the case of $L_y=1$, we can identify widely separated positions where strong coupling occurs. Note that these position states are occupied only by common energy eigenstates near the Fermi level. Focusing on such pairs of $x$ positions, now we consider the case of $L_y>1$.}
%\tcr{Instead of introducing our previous work, emphasize what is the criteria to have exotic interaction which follows in the next subsection, such as discreteness of electronic states in quasiperiodicity and subbands due to stacking }

\textit{\tcm{Immortal interaction---}}
%Before discussing the general stacking cases with $L_y>1$, we first review the single-layer case with $L_y=1$. For large $\kappa$, there exists a set of widely separated position states that are occupied only by common eigenstates near the Fermi level. This occurs because most electronic eigenstates are critical states that are highly localized on a few widely spaced sites. As a result, the localized spins at these sites can strongly interact with each other despite their large separation. Notably, these site pairs can be identified by their local patterns, as the critical wave functions depend on the tiling pattern. With this understanding of $x$ position pairs, we can now proceed to the case of $L_y>1$.
%\tcr{Introduce discussion for a single layer case with citing our previous work} 
%\tcr{better to discuss enough insights of the main physics before showing numerical claculation results?}
Given a pair of $x$ positions, say $x_1$ and $x_2$, that exhibits anomalously strong coupling within each quasiperiodic chain, we analyze $J_{p,q}$ for $p=(x_p,y_p)$ and $q=(x_q,y_q)$, where $x_{p(q)}$ are either $x_1$ or $x_2$ and $0\le y_{p(q)}/a_y< L_y$, where $a_y$ is the unit length of the $y$-axis. Here, we consider the case of $|x_1-x_2|\sim\mathcal{O}(10^3 a)$, where $a$ is the average atomic distance in a Fibonacci chain. As a concrete example, we focus on the half-filled case with $E_F=0$ under the sublattice symmetry. It is important to note that our results can be generalized to the cases with different Fermi levels.

\begin{figure}
  \includegraphics[width=0.48\textwidth]{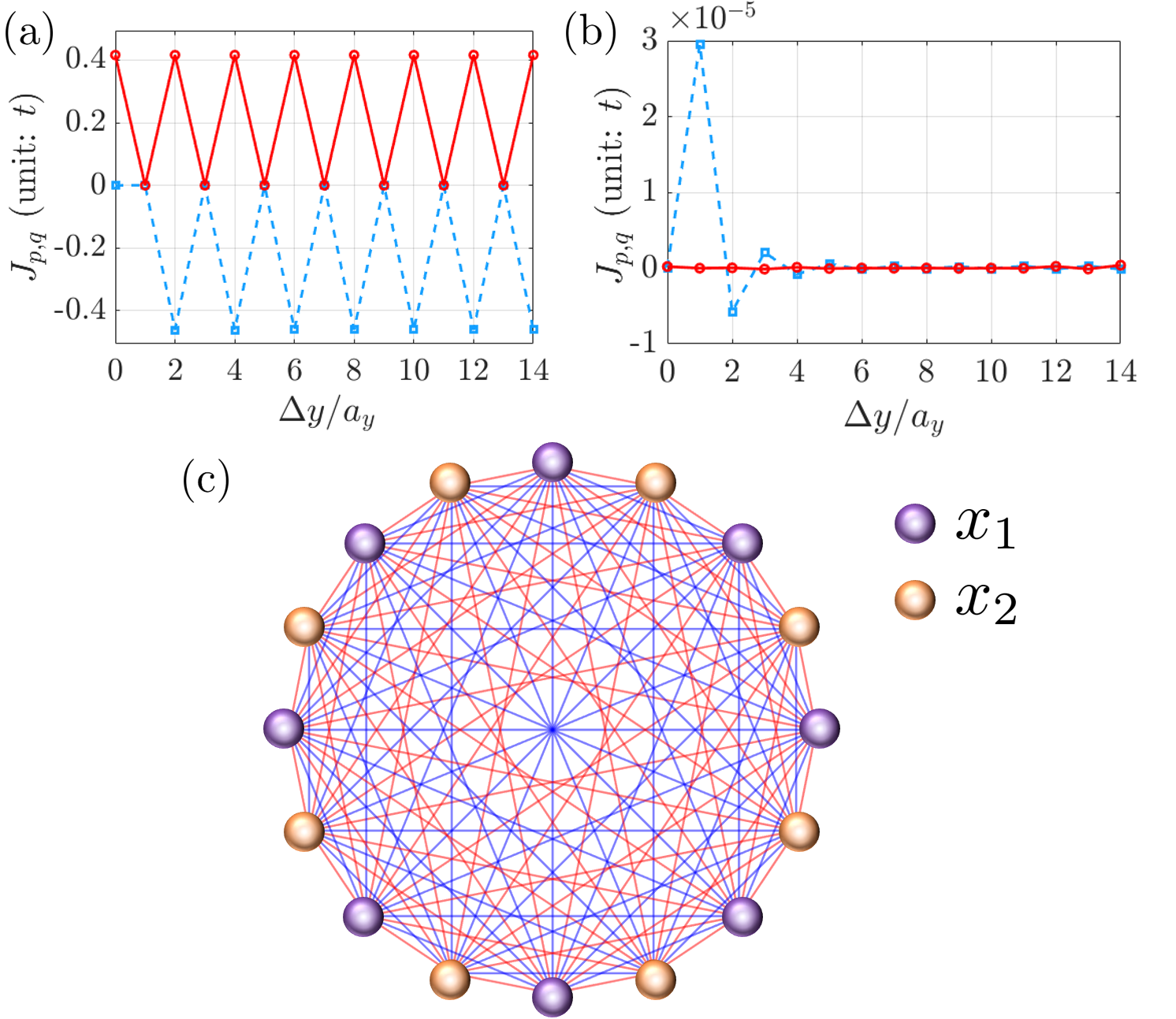}
  \caption{Long-range coupling $J_{p,q}$ as a function of $\Delta y=|y_p-y_q|$ for (a) quasiperiodic case with $\kappa=2.3$ and (b) periodic case with $\kappa=0$, respectively. Red solid lines and blue dashed lines represent the cases of $x_p \neq x_q$ and $x_p = x_q$, respectively. $L_y=15$ and $L_x=1597$. $x_p=305$ and $y_p=0$. $x_q$ is either $305$ or $1292$. $|J_K/t|=0.1$. $\beta^{-1}=10^{-5}t$. (c) Schematic graphical representation of the immortal interaction in (a). Violet and orange spheres represent sites in different sublattices with $x$-coordinates, $x_1$ and $x_2$, respectively. Sites with negligible interaction strength are omitted. Here, red and blue lines represent the couplings shown as the red solid line and the blue dashed line in (a), respectively.}
  \label{fig: result1}
\end{figure}
Fig.\ref{fig: result1} illustrates $J_{p,q}$ for $\kappa=0$ and $\kappa\neq 0$, as a function of $\Delta y/a_y=|y_p-y_q|/a_y$. For the perfectly periodic case with $\kappa=0$, the coupling strength shows conventional attenuation. Thus, $J_{p,q}$ is significant only for the neighboring sites (See Fig.\ref{fig: result1} (b)). On the other hand, for quasiperiodic case with $\kappa \neq 0$, the $J_{p,q}$ interaction is substantial even between widely separated positions and does not diminish along the periodic stacking direction. In detail, $J_{p,q}$ shown in Fig.\ref{fig: result1} (a) oscillates with a period of $2a_y$ while maintaining nearly constant amplitudes, in contrast to the case of $\kappa=0$ depicted in Fig.\ref{fig: result1} (b). It is surprising that the coupling strength does not diminish in the $y$ direction, despite the periodic stacking of the chains. We will refer to this unexpected non-decaying interaction as the \textit{immortal interaction}. 
As a graphical representation of connectivity, Fig.\ref{fig: result1} (c) shows how the immortal interaction, $J_{p,q}$, connects different sites $p$ and $q$, where the solid lines of the same color represent the same $J_{p,q}$. Here, sites with negligible $J_{p,q}$ values are omitted.

%\tcg{It turns out that the positions showing the immortal interaction with a given site are occupied by common eigenstates around the Fermi level, despite being located at different $y$ positions. This implies that the $y$ degrees of freedom is compressed at these positions. To understand how such compression occurs,}
To understand such immortal interaction, we first note that the electronic energy can be expressed as the sum of the energies originating from $T_x$ and $T_y$, respectively, because $[T_x,T_y]=0$. Since the energy levels of $T_x$, which correspond to the $x$-position states of $x_1$ and $x_2$ in a Fibonacci chain, are highly concentrated near the zero energy for large $\kappa$, the energy contribution from $T_y$ must also be zero in order for the total energy $T_x+T_y$ to be near the Fermi level. 
Note that due to the quasi-1D nature, the subbands sparsely occupy the energy spectrum. This effectively compresses the $y$ degrees of freedom to $k_ya_y=\pi/{2}$, resulting in $T_y\ket{k_ya_y=\pi/{2}}=0$ (See Supplementary Materials for detailed information.). Consequently, the immortal interaction oscillates with a period of $2a_y$. In detail, the immortal interaction is nearly zero for an odd $|y_p-y_q|/a_y$, while it is uniformly strong for an even $|y_p-y_q|/a_y$. We note that the immortal interaction occurs only for odd $L_y$, where $T_y$ admits the zero energy.  
Thus, one can conclude the interplay of quasiperiodicity and the subband characteristics of quasi-1D system  
leads to a highly restricted phase space of electronic states in determining the indirect spin interaction. This results in the emergence of a long-range, non-decaying phenomenon, \textit{immortal interaction}.

\textit{\tcm{Eccentric entanglement driven by immortal interactions ---}} The immortal interaction is expected to produce unique entanglement properties. Note that conventional decaying interactions typically lead to area law scaling of entanglement, which is proportional to the boundary area between subsystems\cite{eisert2008area}. Thus, the long-range entanglement decreases rapidly with increasing temperature or distance\cite{gong2017entanglement,eisert2008area}. While, the immortal interaction can modify these conventional properties of entanglement without requiring intermediary qubits, owing to its anomalously enhanced strength and the monogamy of entanglement.

%\tcg{For $J_K>0$, the localized electron spin is screened by forming a local singlet below a critical temperature known as the Kondo temperature\cite{chaikin1995principles}. This phenomenon is known as the Kondo screening\cite{jones1988low}. 
%Meanwhile, when $J_K<0$ or in low-temperature (RKKY) regime, Kondo screening does not occur\cite{zhang2013temperature}. Hence, localized spins act as long-range interacting moments, leading to no entanglement between the two in low-energy states\cite{jones1988low}.}

\begin{figure}
  \includegraphics[width=0.48\textwidth]{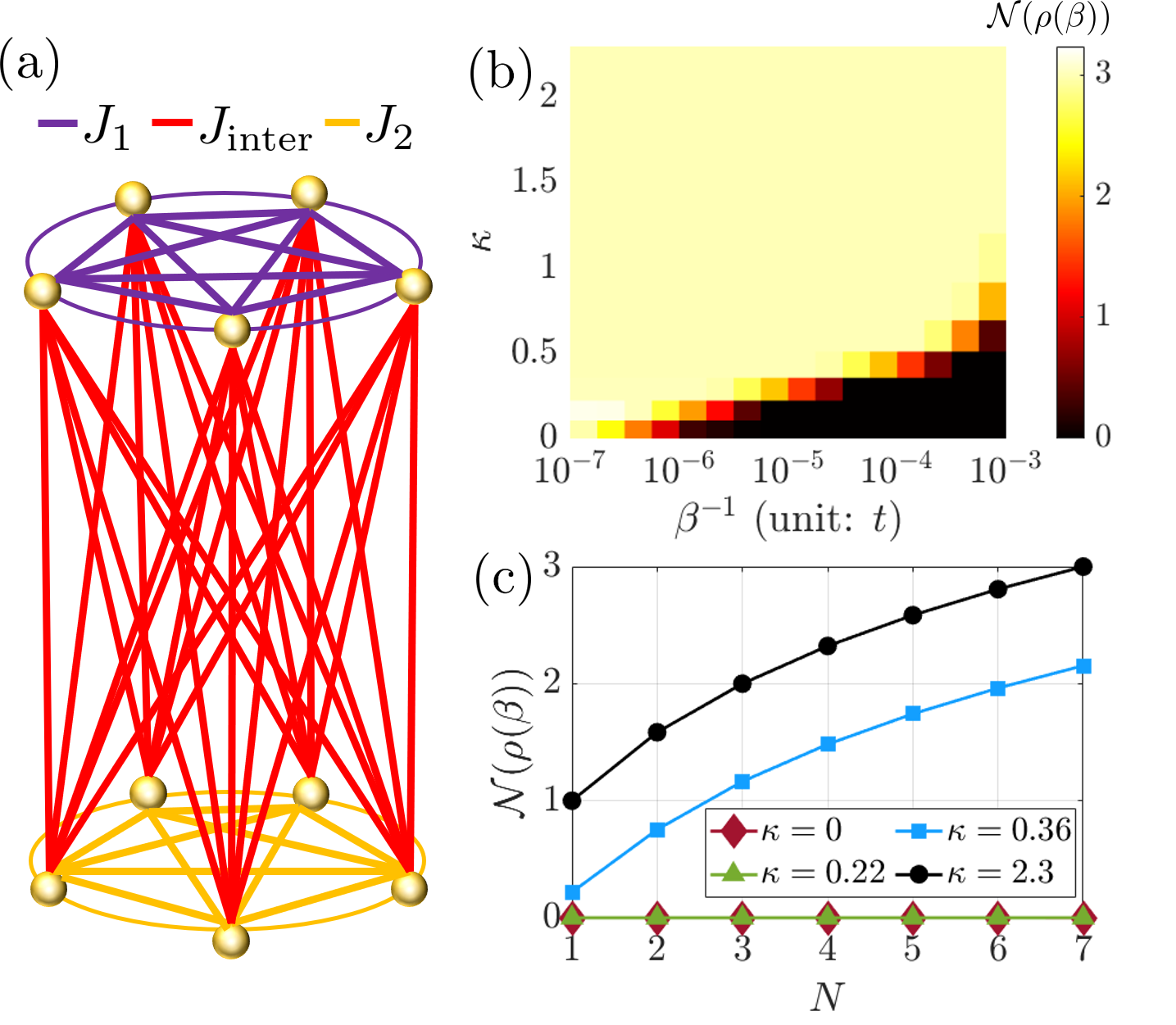}
  \caption{Entanglement characteristics induced by immortal interactions. (a) Schematic graphical representation of the immortal interactions between two parties, highlighted as violet and orange shaded regions in Fig.\ref{fig: system}, exemplified for $N=5$. Violet, orange lines represent the couplings within each party, $J_1, J_2$, respectively, while red lines represent the couplings between parties, $J_{\mathrm{inter}}$. (b) Negativity between two parties as functions of temperature and quasiperiodicity strength for $N=7$. The negativity remains constant at $3$ for $\kappa>0.5$, even at finite temperature $\beta^{-1}=10^{-4}t$. (c) Negativity as a function of the size of each party, $N$ for $\beta^{-1}=10^{-5}t$ and different values of $\kappa$. Black circle and blue square curves show logarithmic scaling. Particularly, the black circle curve is given by $\mathcal{N}(\rho(\beta))=\log_2(N+1)$. Here, $L_x=1597$. $x_1=305$ and $x_2=1292$. $|J_K/t|=0.1$.}
  \label{fig: result2}
\end{figure}

%\tcg{Now let's focus on the weak coupling regime with $J_K<0$ and $\beta<\infty$.} 
To compute the entanglement between localized spins using the thermal density matrix at finite temperature, $\rho(\beta)=e^{-\beta H_{\mathrm{loc}}}/\mathrm{Tr}(e^{-\beta H_{\mathrm{loc}}})$, we adopt negativity as the measure of thermal entanglement. Here, $H_{\mathrm{\mathrm{loc}}}=\sum_{p\neq q}J_{p,q}\vec{S}_p\cdot\vec{S}_q$ is the effective Hamiltonian for the localized spins. We group localized spins into two parties based on their $x$-positions, $x_1$ and $x_2$ that admit antiferromagnetic couplings between parties while ferromagnetic couplings within each party, as shown in Fig.\ref{fig: result1} (a). Two parties, separated by a mesoscopic distance, are represented by violet and orange shaded regions in Fig.\ref{fig: system}. Since the immortal interaction oscillates as a function of $\Delta y$ with a period of $2a_y$, $N=(L_y+1)/2$ localized spins in each party interact via the immortal couplings with a large $\kappa$ and an odd $L_y$. Fig.\ref{fig: result2} (a) provides a graphical representation of how $2N$ spins, belonging to two parties, are coupled through immortal interactions, with an example for $N=5$. Note that there are three different immortal couplings. The negativity of $\rho(\beta)$ is given by $\mathcal{N}(\rho(\beta)) = \log_2(||\rho(\beta)^{\Gamma_1}||)$, where $\Gamma_1$ denotes the partial transpose operation with respect to the $x_1$ subsystem, and $||\cdot||$ represents the trace norm.

Fig.\ref{fig: result2} illustrates the negativity as functions of $\kappa$, $N$ and temperature. First, when the quasiperiodicity is sufficiently strong, entanglement between two widely separated parties can persist even at finite temperatures. Specifically, Fig.\ref{fig: result2} (b) shows that for small $\kappa$, the negativity quickly disappears at low temperatures, whereas for large $\kappa$, the negativity remains robust even at higher temperatures. Hence, strong quasiperiodicity enables us to utilize long-range quantum correlations at higher temperatures.

Next, we find the unusual scaling behavior of negativity under conditions of sufficiently strong quasiperiodicity. Fig.\ref{fig: result2} (c) shows that for sufficiently large values of $\kappa$, the negativity grows logarithmically with $N$, deviating from the conventional area law where it would typically increase linearly with $N$ (See Supplementary Materials for detailed information.)\cite{gong2017entanglement}. In detail, for given $\beta<\infty$, the negativity is either constantly zero (small $\kappa$) or logarithmically grows (large $\kappa$). When $\kappa$ is small, we have $\mathcal{N}(\rho(\beta))=0$ due to the thermalization. Whereas, for large $\kappa$, we find logarithmic scaling given by
\begin{align}
\label{negativityscaling}
    &\mathcal{N}(\rho(\beta))\sim \log_2(N+1).
\end{align} Particularly, for $\kappa\gg0$, $\mathcal{N}(\rho(\beta))=\log_2(N+1)$. Note that for given finite temperature, the immortal couplings with $\kappa\gg0$ are strong compared to $\beta^{-1}$. Thus, the density matrix is approximated by the pure ground state, $\vert\Psi_G\rangle\langle\Psi_G\vert$ of $H_{\mathrm{loc}}$. Here, $H_{\mathrm{loc}}$ can be rewritten in terms of the immortal couplings as,
\begin{align}
    \label{reimpurityH}
    H_{\mathrm{\mathrm{loc}}}(\kappa\gg0)=& (J_1-J_{\mathrm{inter}})\vec{S}_{1,\mathrm{tot}}^2+(J_2-J_{\mathrm{inter}})\vec{S}_{2,\mathrm{tot}}^2\\ &+J_{\mathrm{inter}}\vec{S}_\mathrm{tot}^2+\mathrm{const},\nonumber
\end{align}
where $J_1,J_2<0$ are the ferromagnetic immortal coupling within each parties, and $J_{\mathrm{inter}}>0$ is the antiferromagnetic immortal coupling between parties. $\vec{S}_{1,\mathrm{tot}}$, $\vec{S}_{2,\mathrm{tot}}$ are total spins of each party, and $\vec{S}_\mathrm{tot}=\vec{S}_{1,\mathrm{tot}}+\vec{S}_{2,\mathrm{tot}}$. Now, $H_{\mathrm{loc}}$ possesses SU(2) symmetry, and the four operators $H_\mathrm{loc},\vec{S}_{\mathrm{tot}}^2,\vec{S}_{1,\mathrm{tot}}^2$ and $\vec{S}_{2,\mathrm{tot}}^2$ are all commute with each other. Hence, $\vert\Psi_G\rangle=\vert S_1=N/2, S_2=N/2, S=0\rangle$, where $S_1,S_2$ and $S$ are total spin quantum numbers defined by $\vec{S}_{1,\mathrm{tot}}^2,\vec{S}_{2,\mathrm{tot}}^2$ and $\vec{S}_\mathrm{tot}^2$, respectively. Thus, the negativity is given by $\mathcal{N}(\vert\Psi_G\rangle\langle\Psi_G\vert)=\log_2(N+1)$.

The logarithmic scaling of negativity is a consequence of the monogamy of entanglement induced by the immortal interactions\cite{bai2014general}. Since the immortal interaction with sufficiently large $\kappa$ remains constant regardless of the $y$-distances, $N$ spins in one party compete equally to form singlets with spins in the other party. This causes a significant enhancement in the monogamy of entanglement, which in turn suppresses the growth of entanglement. As a result, the scaling behavior of negativity follows a logarithmic pattern as Eq.\eqref{negativityscaling}, deviating from the conventional area law. Therefore, the immortal interaction offers a novel approach to realizing an unconventional phase of matter, characterized by this distinct entanglement scaling behavior.

\textit{\tcm{Conclusion---}} 
In summary, we propose quasiperiodic quasi-1D systems as a novel platform for realizing anomalous long-range interactions. Using periodically stacked Fibonacci chains as an example, the indirect interaction between two spins, even when separated by mesoscopic or macroscopic distances, is shown to be significantly enhanced and, more strikingly, does not decay along the stacking direction — a phenomenon referred to as the \textit{immortal interaction}.
%\tcg{and entanglement emergent in the periodically stacked quasiperiodic chains are clarified. The indirect long-range interaction is not only enhanced for a long-distance due to the quasiperiodicity but also not attenuating along the periodically stacked direction, which we named immortal interaction. The immortal interaction gives rise to the unconventional behaviors of long-range entanglement.}
First, the long-range entanglement is stable with respect to unavoidable thermal fluctuation effect. Hence, such systems would be a key player to develop a stable platform of a wide range of quantum engineering such as quantum teleportation and quantum computing. Second, the entanglement induced by the immortal interaction grows logarithmically since it enhances the effect of the monogamy of entanglement. Thus, the immortal interaction also enables us to reach novel long-range entangled phase of matters\cite{lee2013entanglement,kuwahara2020area,PhysRevX.14.021040}.
To measure this long-range entanglement, negativity is one example as we suggested, which can be quantified through various quantum state tomography techniques\cite{PhysRevLett.109.120403}. Beyond negativity, there are also other experimentally accessible indirect entanglement measures, like quantum Fisher information\cite{sabharwal2024witnessing,scheie2021witnessing}, which can be also applicable to our systems.
%\tcg{In particular, QFI, which can be derived from site-resolved measurements of spin-spin correlations using spin-polarized scanning tunneling microscopy or quantum gas microscopy, serves as a sensitive probe for detecting quantum phase transitions\cite{cheuk2016observation,parsons2016site,wolf2024efficient,christakis2023probing}.}

%\tcg{Periodically stacked quasiperiodic chain structures are found in alloys like Al-Ni-Co\cite{ferralis2005structural,tsuda1996structure,theis2008epitaxial}, and can also be engineered using photonic crystals or cold atom arrays\cite{vardeny2013optics,poddubny2010photonic,corcovilos2019two,singh2015fibonacci}. Heterostructures, such as van der Waals materials and twisted bilayer graphene, can also exhibit similar long-range correlations\cite{nimbalkar2020opportunities,alferov2001nobel,chen2019interface,ahn2018dirac,hofmann2022fermionic,lisi2021observation,choi2019electronic,cao2020tunable}. Hence, investigating more experimentally accessible platforms would be an interesting direction for future research. Additionally, the strong coupling regime with antiferromagnetic Kondo interactions offers another intriguing area of study.}
Our findings can be applied to a variety of distinct systems, such as quasiperiodic alloy materials \cite{theis2008epitaxial,jagannathan2021fibonacci}, artificially engineered photonic crystals and cold atom arrays\cite{corcovilos2019two,singh2015fibonacci}.
Moreover, our physics extend beyond the current scope, offering potential applications even in quasiperiodic quasi-2D systems, such as heterostructures and multilayer van der Waals materials with twisting angles that induce quasiperiodicity\cite{alferov2001nobel,cao2020tunable}. These discoveries pave the way for designing exotic quantum phases, stronger long-range interactions, and robust quantum entanglement in quantum materials.

%In the strong coupling regime, the Kondo screening effect compete with the long-range interaction. This would be expected to open the new phases of matter. For instance, one can imagine the exotic phase where local Kondo screened singlet and long-range entangled pairs coexist depending on the two local energy scales, Kondo temperature and long-range coupling.

\section*{Acknowledgments}%\nonumber

%We thank Takanori Sugimoto and Taku J Sato for useful discussions.
J.M.J and S.B.L. were supported by National Research Foundation Grant (2021R1A2C109306013) and Nano Material Technology Development Program through the National Research Foundation of Korea(NRF) funded by Ministry of Science and ICT (RS-2023-00281839). This research was supported in part by grant NSF PHY-1748958 to the Kavli Institute for Theoretical Physics (KITP).

\bibliography{my1}

\begin{thebibliography}{10}

\bibitem{lars2018second}
J~Lars.
\newblock {\em The second quantum revolution: From entanglement to quantum
  computing and other super-technologies}.
\newblock Springer International Publishing, 2018.

\bibitem{hu2023progress}
Xiao-Min Hu, Yu~Guo, Bi-Heng Liu, Chuan-Feng Li, and Guang-Can Guo.
\newblock Progress in quantum teleportation.
\newblock {\em Nature Reviews Physics}, 5(6):339--353, 2023.

\bibitem{cai2015entanglement}
X-D Cai, Dian Wu, Z-E Su, M-C Chen, X-L Wang, Li~Li, N-L Liu, C-Y Lu, and J-W
  Pan.
\newblock Entanglement-based machine learning on a quantum computer.
\newblock {\em Physical review letters}, 114(11):110504, 2015.

\bibitem{zeng2015quantum}
Bei Zeng, Xie Chen, Duan-Lu Zhou, and Xiao-Gang Wen.
\newblock Quantum information meets quantum matter--from quantum entanglement
  to topological phase in many-body systems.
\newblock {\em arXiv preprint arXiv:1508.02595}, 2015.

\bibitem{cramer2011measuring}
Marcus Cramer, Martin~B Plenio, and Harald Wunderlich.
\newblock Measuring entanglement in condensed matter systems.
\newblock {\em Physical review letters}, 106(2):020401, 2011.

\bibitem{hamma2008entanglement}
A~Hamma, W~Zhang, S~Haas, and DA~Lidar.
\newblock Entanglement, fidelity, and topological entropy in a quantum phase
  transition to topological order.
\newblock {\em Physical Review B—Condensed Matter and Materials Physics},
  77(15):155111, 2008.

\bibitem{togan2010quantum}
Emre Togan, Yiwen Chu, Alexei~S Trifonov, Liang Jiang, Jeronimo Maze, Lilian
  Childress, MV~Gurudev Dutt, Anders~S{\o}ndberg S{\o}rensen, Phillip~R Hemmer,
  Alexander~S Zibrov, et~al.
\newblock Quantum entanglement between an optical photon and a solid-state spin
  qubit.
\newblock {\em Nature}, 466(7307):730--734, 2010.

\bibitem{fiaschi2021optomechanical}
Niccol{\`o} Fiaschi, Bas Hensen, Andreas Wallucks, Rodrigo Benevides, Jie Li,
  Thiago P~Mayer Alegre, and Simon Gr{\"o}blacher.
\newblock Optomechanical quantum teleportation.
\newblock {\em Nature Photonics}, 15(11):817--821, 2021.

\bibitem{yin2012quantum}
Juan Yin, Ji-Gang Ren, He~Lu, Yuan Cao, Hai-Lin Yong, Yu-Ping Wu, Chang Liu,
  Sheng-Kai Liao, Fei Zhou, Yan Jiang, et~al.
\newblock Quantum teleportation and entanglement distribution over
  100-kilometre free-space channels.
\newblock {\em Nature}, 488(7410):185--188, 2012.

\bibitem{agarwal2023long}
Lakshya Agarwal, Christopher~M Langlett, and Shenglong Xu.
\newblock Long-range bell states from local measurements and many-body
  teleportation without time reversal.
\newblock {\em Physical Review Letters}, 130(2):020801, 2023.

\bibitem{osterloh2002scaling}
Andreas Osterloh, Luigi Amico, Giuseppe Falci, and Rosario Fazio.
\newblock Scaling of entanglement close to a quantum phase transition.
\newblock {\em Nature}, 416(6881):608--610, 2002.

\bibitem{savary2016quantum}
Lucile Savary and Leon Balents.
\newblock Quantum spin liquids: a review.
\newblock {\em Reports on Progress in Physics}, 80(1):016502, 2016.

\bibitem{isakov2011topological}
Sergei~V Isakov, Matthew~B Hastings, and Roger~G Melko.
\newblock Topological entanglement entropy of a bose--hubbard spin liquid.
\newblock {\em Nature Physics}, 7(10):772--775, 2011.

\bibitem{lee2013entanglement}
Yirun~Arthur Lee and Guifre Vidal.
\newblock Entanglement negativity and topological order.
\newblock {\em Physical Review A—Atomic, Molecular, and Optical Physics},
  88(4):042318, 2013.

\bibitem{sabharwal2024witnessing}
Snigdh Sabharwal, Tokuro Shimokawa, and Nic Shannon.
\newblock Witnessing disorder in quantum magnets.
\newblock {\em arXiv preprint arXiv:2407.20797}, 2024.

\bibitem{scheie2021witnessing}
Allen Scheie, Pontus Laurell, AM~Samarakoon, B~Lake, SE~Nagler, GE~Granroth,
  S~Okamoto, G~Alvarez, and DA~Tennant.
\newblock Witnessing entanglement in quantum magnets using neutron scattering.
\newblock {\em Physical Review B}, 103(22):224434, 2021.

\bibitem{gabbrielli2018multipartite}
Marco Gabbrielli, Augusto Smerzi, and Luca Pezz{\`e}.
\newblock Multipartite entanglement at finite temperature.
\newblock {\em Scientific reports}, 8(1):15663, 2018.

\bibitem{gong2017entanglement}
Zhe-Xuan Gong, Michael Foss-Feig, Fernando~GSL Brand{\~a}o, and Alexey~V
  Gorshkov.
\newblock Entanglement area laws for long-range interacting systems.
\newblock {\em Physical review letters}, 119(5):050501, 2017.

\bibitem{defenu2023long}
Nicolo Defenu, Tobias Donner, Tommaso Macr{\`\i}, Guido Pagano, Stefano Ruffo,
  and Andrea Trombettoni.
\newblock Long-range interacting quantum systems.
\newblock {\em Reviews of Modern Physics}, 95(3):035002, 2023.

\bibitem{degli2006long}
C~Degli Esposti~Boschi, M~Roncaglia, et~al.
\newblock Long-distance entanglement in spin systems.
\newblock {\em Physical Review Letters}, 96(24):247206--247206, 2006.

\bibitem{jin2004localizable}
B-Q Jin and VE~Korepin.
\newblock Localizable entanglement in antiferromagnetic spin chains.
\newblock {\em Physical Review A—Atomic, Molecular, and Optical Physics},
  69(6):062314, 2004.

\bibitem{heiss2008fundamentals}
Dieter Heiss.
\newblock {\em Fundamentals of quantum information: quantum computation,
  communication, decoherence and all that}, volume 587.
\newblock Springer, 2008.

\bibitem{eisert2008area}
Jens Eisert, Marcus Cramer, and Martin~B Plenio.
\newblock Area laws for the entanglement entropy-a review.
\newblock {\em arXiv preprint arXiv:0808.3773}, 2008.

\bibitem{eisert2006general}
Jens Eisert and Tobias~J Osborne.
\newblock General entanglement scaling laws from time evolution.
\newblock {\em Physical review letters}, 97(15):150404, 2006.

\bibitem{fannes2003entropy}
Mark Fannes, Bart Haegeman, and M~Mosonyi.
\newblock Entropy growth of shift-invariant states on a quantum spin chain.
\newblock {\em Journal of Mathematical Physics}, 44(12):6005--6019, 2003.

\bibitem{arguello2022tuning}
Javier Arg{\"u}ello-Luengo, Alejandro Gonz{\'a}lez-Tudela, and Daniel
  Gonz{\'a}lez-Cuadra.
\newblock Tuning long-range fermion-mediated interactions in cold-atom quantum
  simulators.
\newblock {\em Physical review letters}, 129(8):083401, 2022.

\bibitem{saffman2010quantum}
Mark Saffman, Thad~G Walker, and Klaus M{\o}lmer.
\newblock Quantum information with rydberg atoms.
\newblock {\em Reviews of modern physics}, 82(3):2313--2363, 2010.

\bibitem{estrada2024correlation}
Juan~Carlos Estrada~Salda{\~n}a, Alexandros Vekris, Luka Pave{\v{s}}i{\v{c}},
  Rok {\v{Z}}itko, Kasper Grove-Rasmussen, and Jesper Nyg{\aa}rd.
\newblock Correlation between two distant quasiparticles in separate
  superconducting islands mediated by a single spin.
\newblock {\em Nature Communications}, 15(1):3465, 2024.

\bibitem{lahaye2009physics}
Thierry Lahaye, C~Menotti, L~Santos, M~Lewenstein, and T~Pfau.
\newblock The physics of dipolar bosonic quantum gases.
\newblock {\em Reports on Progress in Physics}, 72(12):126401, 2009.

\bibitem{islam2013emergence}
R~Islam, Crystal Senko, Wes~C Campbell, S~Korenblit, J~Smith, A~Lee,
  EE~Edwards, C-CJ Wang, JK~Freericks, and C~Monroe.
\newblock Emergence and frustration of magnetism with variable-range
  interactions in a quantum simulator.
\newblock {\em science}, 340(6132):583--587, 2013.

\bibitem{liao2022end}
Ying-Yen Liao.
\newblock End-to-end entanglement in a polar-molecule array under intrinsic
  decoherence.
\newblock {\em Physical Review A}, 105(6):062802, 2022.

\bibitem{jacquod2009decoherence}
Ph~Jacquod and Cyril Petitjean.
\newblock Decoherence, entanglement and irreversibility in quantum dynamical
  systems with few degrees of freedom.
\newblock {\em Advances in Physics}, 58(2):67--196, 2009.

\bibitem{buchleitner2008entanglement}
Andreas Buchleitner, Carlos Viviescas, and Markus Tiersch.
\newblock {\em Entanglement and decoherence: foundations and modern trends},
  volume 768.
\newblock Springer Science \& Business Media, 2008.

\bibitem{jagannathan2021fibonacci}
Anuradha Jagannathan.
\newblock The fibonacci quasicrystal: Case study of hidden dimensions and
  multifractality.
\newblock {\em Reviews of Modern Physics}, 93(4):045001, 2021.

\bibitem{jeon2023anomalous}
Junmo Jeon and SungBin Lee.
\newblock Anomalous long-distance rkky interaction in quasicrystals.
\newblock {\em arXiv preprint arXiv:2310.15228}, 2023.

\bibitem{chaikin1995principles}
Paul~M Chaikin, Tom~C Lubensky, and Thomas~A Witten.
\newblock {\em Principles of condensed matter physics}, volume~10.
\newblock Cambridge university press Cambridge, 1995.

\bibitem{bai2014general}
Yan-Kui Bai, Yuan-Fei Xu, and ZD~Wang.
\newblock General monogamy relation for the entanglement of formation in
  multiqubit systems.
\newblock {\em Physical review letters}, 113(10):100503, 2014.

\bibitem{kuwahara2020area}
Tomotaka Kuwahara and Keiji Saito.
\newblock Area law of noncritical ground states in 1d long-range interacting
  systems.
\newblock {\em Nature communications}, 11(1):4478, 2020.

\bibitem{PhysRevX.14.021040}
Nathanan Tantivasadakarn, Ryan Thorngren, Ashvin Vishwanath, and Ruben
  Verresen.
\newblock Long-range entanglement from measuring symmetry-protected topological
  phases.
\newblock {\em Phys. Rev. X}, 14:021040, Jun 2024.

\bibitem{PhysRevLett.109.120403}
Matthias Christandl and Renato Renner.
\newblock Reliable quantum state tomography.
\newblock {\em Phys. Rev. Lett.}, 109:120403, Sep 2012.

\bibitem{theis2008epitaxial}
Wolfgang Theis and KJ~Franke.
\newblock Epitaxial interfaces between half-crystals of quasicrystalline and
  periodic material.
\newblock {\em Journal of Physics: Condensed Matter}, 20(31):314004, 2008.

\bibitem{corcovilos2019two}
Theodore~A Corcovilos and Jahnavee Mittal.
\newblock Two-dimensional optical quasicrystal potentials for ultracold atom
  experiments.
\newblock {\em Applied optics}, 58(9):2256--2263, 2019.

\bibitem{singh2015fibonacci}
Kevin Singh, Kush Saha, Siddharth~A Parameswaran, and David~M Weld.
\newblock Fibonacci optical lattices for tunable quantum quasicrystals.
\newblock {\em Physical Review A}, 92(6):063426, 2015.

\bibitem{alferov2001nobel}
Zhores~I Alferov.
\newblock Nobel lecture: The double heterostructure concept and its
  applications in physics, electronics, and technology.
\newblock {\em Reviews of modern physics}, 73(3):767, 2001.

\bibitem{cao2020tunable}
Yuan Cao, Daniel Rodan-Legrain, Oriol Rubies-Bigorda, Jeong~Min Park, Kenji
  Watanabe, Takashi Taniguchi, and Pablo Jarillo-Herrero.
\newblock Tunable correlated states and spin-polarized phases in twisted
  bilayer--bilayer graphene.
\newblock {\em Nature}, 583(7815):215--220, 2020.

\end{thebibliography}
\bibliographystyle{unsrt}

\newpage

\renewcommand{\thesection}{\arabic{section}}
\setcounter{section}{0}
\renewcommand{\thefigure}{S\arabic{figure}}
\setcounter{figure}{0}
\renewcommand{\theequation}{S\arabic{equation}}
\setcounter{equation}{0}

\begin{widetext}
	\section*{Supplementary Materials for Immortal quantum correlation in quasiperiodic quasi-1D system}
%\section{Supplementary Figures}
\section{Subbands nature of quasi-1D system}
    \label{sec:0}
    In this section, we illustrate the subband nature of a quasi-1D system that leads to immortal interactions. Fig.\ref{fig: supp0} presents the subband structures for various strengths of quasiperiodicity. Recall that the states near the Fermi level, $E_F=0$, primarily contribute to the long-range indirect interaction. As the quasiperiodicity strength $\kappa$ increases, the states near the zero energy appear exclusively at $k_ya_y=\pi/{2}$, where $a_y$ is the unit length along the $y$-axis. This indicates a compression of the $y$ degrees of freedom to $k_ya_y=\pi/{2}$, resulting in immortal couplings.
    \begin{figure}[h]
    \includegraphics[width=0.9\textwidth]{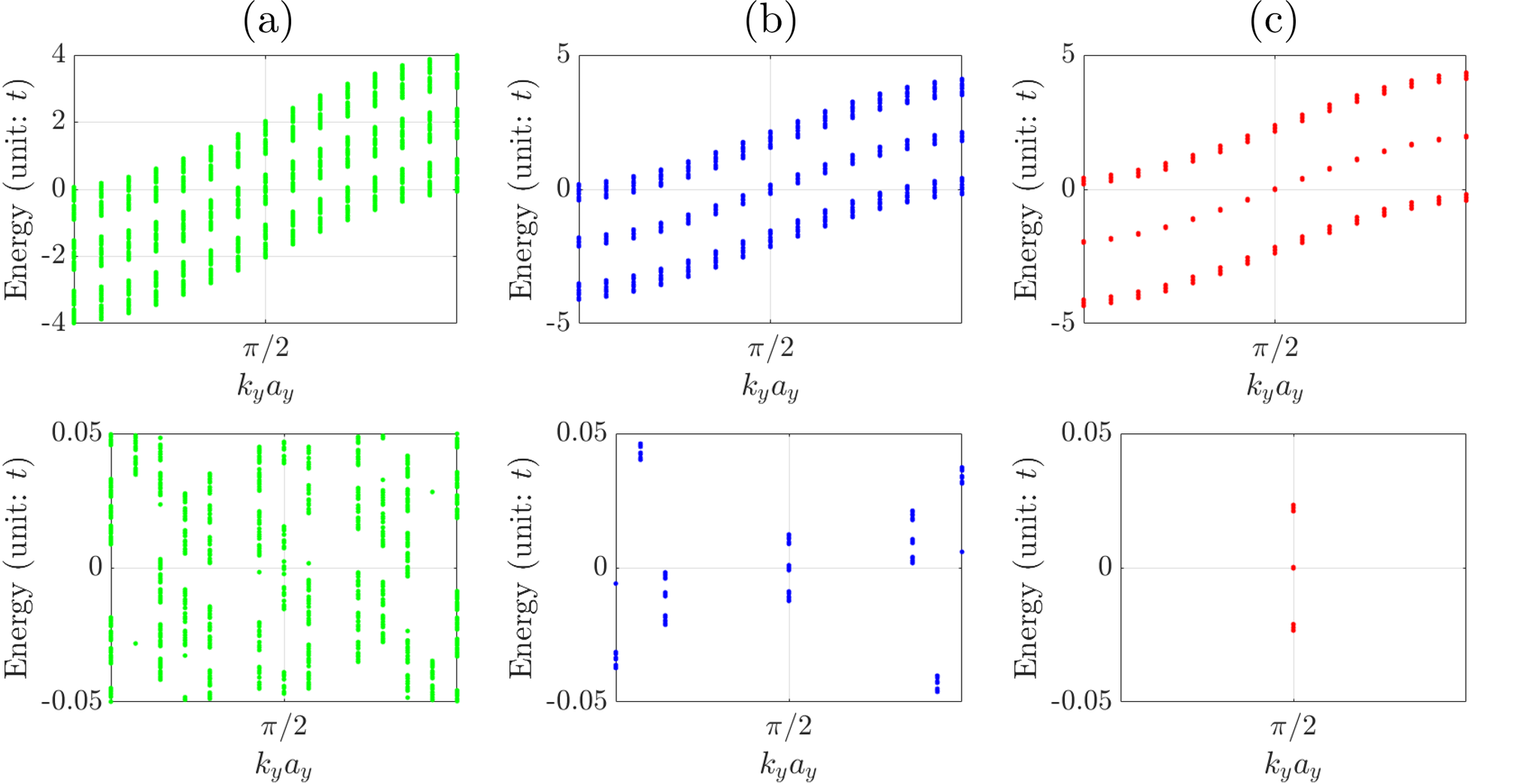}
  \caption{Subband structures for different strengths of quasiperiodicity: (a) $\kappa=0.22$, (b) $\kappa=0.52$, (c) $\kappa=2.3$. The bottom panels show a zoomed-in view near zero energy. For sufficiently large $\kappa$, the states near the zero energy appear exclusively at $k_ya_y = \pi/{2}$. Here, $L_x=1597$ and $L_y=15$.}
  \label{fig: supp0}
\end{figure}

\section{Area law scaling behavior of entanglement in the absence of quasiperiodicity}
	\label{sec:0}
Short-range or rapidly decaying long-range interactions typically result in area law scaling of entanglement, unless the subsystems are uncorrelated. This is because these types of interactions concentrate mutual information primarily at the boundaries between subsystems. In this section, we demonstrate the standard area law scaling of entanglement by examining two adjacent subsystems in the absence of quasiperiodicity ($\kappa=0$), represented by the violet and orange shaded regions in Fig. \ref{fig: supp1}(a). The coupling strength, $J_{p,q}$ between these subsystems decreases uniformly with distance, as illustrated in Fig. \ref{fig: supp1}(b).

%In this section, we aim to demonstrate the standard area law scaling of entanglement for the case of $\kappa=0$, where the coupling strength uniformly decays with distance (see Fig.\ref{fig: supp1} (b)). To obtain nonzero quantum correlations, we now focus on two neighboring subsystems, depicted as the violet and orange shaded regions in Fig.\ref{fig: supp1} (a).%See Fig.\ref{fig: supp1} (b) for the indirect exchange couplings, $J_{p,q}$, between two neighboring subsystems, which decrease uniformly with distance.

Fig.\ref{fig: supp1} (c) shows the linear increase in negativity between two neighboring subsystems for $\kappa=0$ at finite temperature. It is important to note that the boundary area between the subsystems increases linearly with $N$. This indicates that the entanglement between neighboring subsystems follows the standard area law scaling behavior in the absence of quasiperiodicity.
\begin{figure}[h]
    \includegraphics[width=0.9\textwidth]{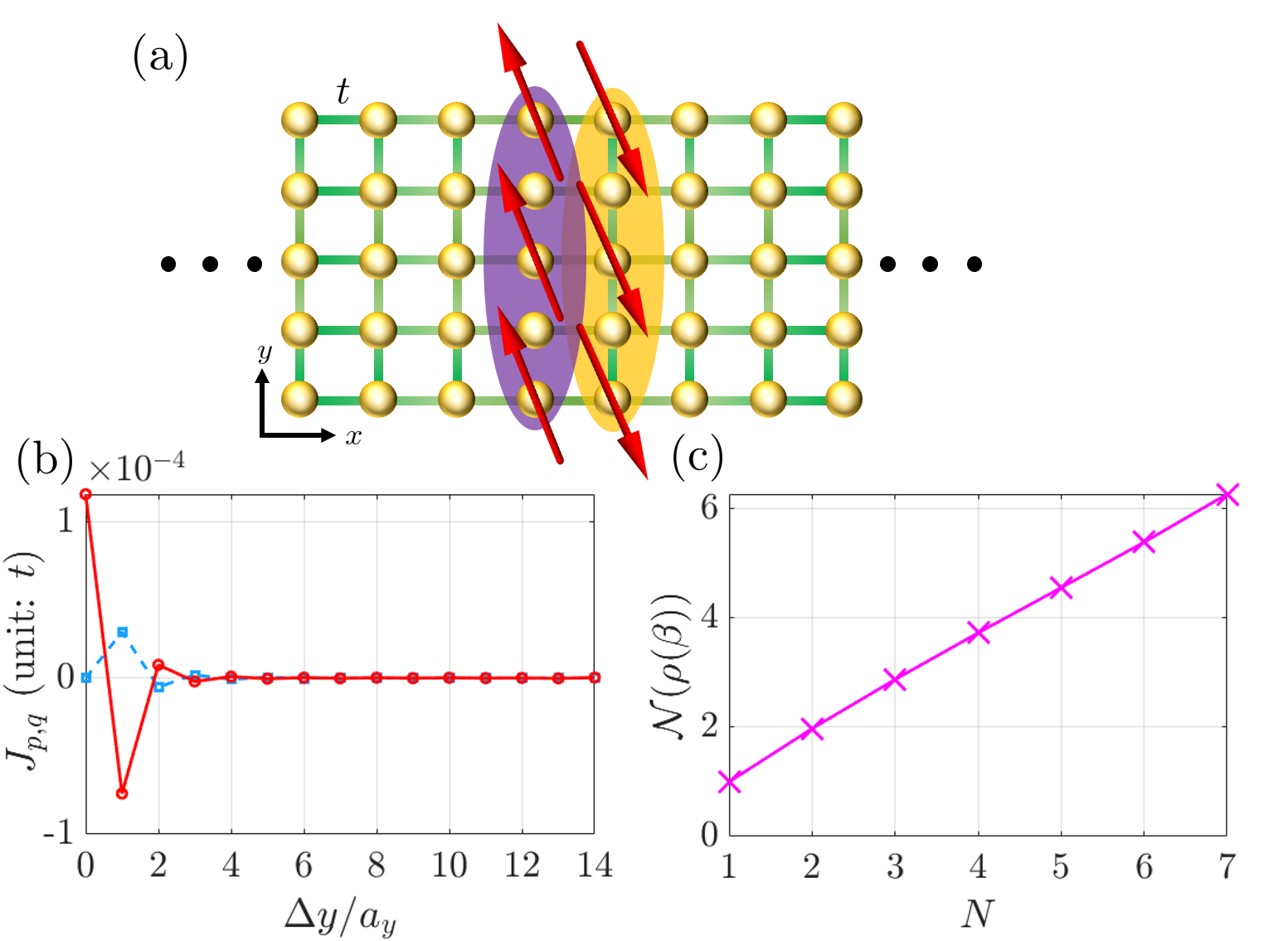}
  \caption{(a) Schematic illustration of two neighboring subsystems in a periodic quasi-1D system with $\kappa=0$. Green bars represent uniform hopping integrals, $t$, while red arrows and yellow spheres depict localized spins and sites, respectively. The violet and orange shaded regions highlight the two neighboring subsystems. The boundary area between the subsystems increases linearly with the subsystem size, $N$. (b) Long-range coupling $J_{p,q}$ as a function of $\Delta y=\vert y_p-y_q\vert$. Red solid lines and blue dashed lines represent the cases of $x_p\neq x_q$ and $x_p=x_q$, respectively. $L_y=15$ and $L_x=1597$. $x_p=305$ and $y_p=0$. $x_q$ is either 305 or 306. $\vert J_K/t\vert=0.1$ and $\beta^{-1}=10^{-5}t$. (c) Negativity between two neighboring subsystems with $\kappa=0$ at finite temperature ($\beta^{-1}=10^{-5}t$) as a function of the subsystem size $N$. The linear growth of negativity demonstrates the conventional area law scaling behavior in the absence of quasiperiodicity. Each subsystem is defined by their $x$-positions: $x_1=305$ and $x_2=306$, respectively.}
  \label{fig: supp1}
\end{figure}
\end{widetext}

\end{document}